\documentclass[12pt]{article}
\usepackage{a4wide,epsfig}

\newcommand{\lmu}[1]{L_{#1}}
\newcommand{\sss}{S(S+1)}
\newcommand{\LrmuEuler}{L_r}
\newcommand{\Las}{L_{\alpha_s}}
\newcommand{\LE}[1]{L^E_{#1}}
\newcommand{\PolyEuler}[1]{P_{#1}}

\catcode`@=11
\newcount\@tempcntc
\def\@citex[#1]#2{\if@filesw\immediate\write\@auxout{\string\citation{#2}}\fi
  \@tempcnta\z@\@tempcntb\m@ne\def\@citea{}\@cite{\@for\@citeb:=#2\do
    {\@ifundefined
       {b@\@citeb}{\@citeo\@tempcntb\m@ne\@citea\def\@citea{,}{\bf ?}\@warning
       {Citation `\@citeb' on page \thepage \space undefined}}%
    {\setbox\z@\hbox{\global\@tempcntc0\csname b@\@citeb\endcsname\relax}%
     \ifnum\@tempcntc=\z@ \@citeo\@tempcntb\m@ne
       \@citea\def\@citea{,}\hbox{\csname b@\@citeb\endcsname}%
     \else
      \advance\@tempcntb\@ne
      \ifnum\@tempcntb=\@tempcntc
      \else\advance\@tempcntb\m@ne\@citeo
      \@tempcnta\@tempcntc\@tempcntb\@tempcntc\fi\fi}}\@citeo}{#1}}
\def\@citeo{\ifnum\@tempcnta>\@tempcntb\else\@citea\def\@citea{,}%
  \ifnum\@tempcnta=\@tempcntb\the\@tempcnta\else
   {\advance\@tempcnta\@ne\ifnum\@tempcnta=\@tempcntb \else \def\@citea{--}\fi
    \advance\@tempcnta\m@ne\the\@tempcnta\@citea\the\@tempcntb}\fi\fi}
\catcode`@=12

\begin{document}


\title{\vskip-3cm{\baselineskip14pt
    \begin{flushleft}
      \normalsize TTP/05-01 \\
      \normalsize SFB/CPP-05-01 \\
      \normalsize DESY 05-001
  \end{flushleft}}
  \vskip1.5cm
  Heavy Quarkonium Spectrum and Production/Annihilation Rates
  to order $\beta_0^3\alpha_s^3$}
\author{\small A.A. Penin$^{a,b}$,
  V.A. Smirnov$^{c,d}$, M. Steinhauser$^a$\\
  {\small\it $^a$ Institut f{\"u}r Theoretische Teilchenphysik,
    Universit{\"a}t Karlsruhe,}\\
  {\small\it 76128 Karlsruhe, Germany}\\
  {\small\it $^b$ Institute for Nuclear Research, 
    Russian Academy of Sciences,}\\
  {\small\it 117312 Moscow, Russia}\\
  {\small\it $^c$ Institute for Nuclear Physics, 
    Moscow State University,}\\
  {\small\it 119992 Moscow, Russia}\\
  {\small\it $^d$ II. Institut f\"ur Theoretische Physik, 
    Universit\"at Hamburg,}\\
  {\small\it 22761 Hamburg, Germany}
}

\date{}

\maketitle

\thispagestyle{empty}

\begin{abstract}
  We compute the third-order corrections to the heavy quarkonium
  spectrum and production/annihilation rates due to the leading
  renormalization group running of the static potential.  The previously
  known complete ${\cal O}(m_q\alpha_s^5)$ result for the heavy
  quarkonium ground state energy is extended to the exited states.
  After including the ${\cal O}(\alpha_s^3)$ corrections the
  perturbative results are in surprisingly good agreement with the
  experimental data on the masses of the excited $\Upsilon$ resonances
  and the leptonic width of the $\Upsilon(1S)$ meson. The impact of the
  corrections on the $\Upsilon$ sum rules and top quark-antiquark
  threshold production cross section is also discussed.

\medskip

\noindent
PACS numbers: 12.38.Bx, 14.40.Gx, 14.65.Ha

\end{abstract}

\newpage


\section{Introduction}

The theoretical study of nonrelativistic heavy quark-antiquark systems
is among the earliest applications of perturbative quantum
chromodynamics (QCD) \cite{AppPol}.  Its applications to bottomonium
\cite{NSVZ} and top-antitop \cite{FadKho} physics entirely rely on the
first principles of QCD.  In general perturbation theory can be
applied for the analysis of these systems.
Nonperturbative effects \cite{Vol,Leu} are well
under control for the top-antitop system and, at least within the
sum-rule approach, also for bottomonium.  This makes
heavy quark-antiquark systems an ideal laboratory to determine
fundamental parameters of QCD, such as the strong coupling constant
$\alpha_s$ and the heavy-quark masses $m_q$.

The binding energy of the heavy quarkonium state and the value of its
wave function at the origin are among the characteristics of the
heavy-quarkonium system that are of primary phenomenological interest.
The former determines the mass of the bound state resonance, while the
latter controls its production and annihilation rates.

Recently, the heavy quarkonium ground state energy has been computed
through ${\cal O}(\alpha_s^5m_q)$ including the third-order correction
to the Coulomb approximation \cite{PenSte}. The result has been used to
extract $m_b$ from the $\Upsilon(1S)$ meson mass. The properties of the
excited states are more sensitive to the nonperturbative phenomena, and
the corresponding perturbative estimates cannot be used, {\it e.g.}, for
the accurate determination of the heavy-quark mass by direct comparison to
the meson masses.  However, they have to be taken into account in the
framework of the nonrelativistic sum rules \cite{NSVZ} which is based on
the concept of quark-hadron duality and keeps the nonperturbative
effects under control.  Moreover, by investigating the excited states
with reliable perturbative results at hand one can test the effects and
structure of the nonperturbative QCD vacuum. Still only a few states
with small principal quantum numbers $n$ and zero orbital momentum $l$
are of practical interest.

As far as the wave function at the origin is concerned 
a complete result is only
available through ${\cal O}(\alpha_s^2)$ \cite{PenPiv1,MelYel1}.  The
${\cal O}(\alpha_s^2)$ correction has turned out to be so sizeable that
the feasibility of an accurate perturbative analysis was challenged
\cite{gang}, and it appears indispensable to gain full control over the
next order. Only the logarithmically enhanced 
${\cal O}(\alpha_s^3\ln^2\alpha_s)$~\cite{KniPen2,ManSte} and 
${\cal O}(\alpha_s^3\ln\alpha_s)$~\cite{KPSS2,Hoa2}
(for QED, see Refs.~\cite{KniPen3,HilLep,MelYel2})
corrections are available so far.

In this paper, we take the next step and calculate the nonlogarithmic
third-order corrections to the wave function at the origin and to the
spectrum of the excited heavy quarkonium states proportional to
$\beta_0^3$, where $\beta_0$ is the one-loop QCD beta-function.
Together with the contributions already known, the new term allows to
derive the complete result for the binding energy of the excited states.
On the other hand the large-$\beta_0$ terms usually constitute a
considerable part of the corrections and can be used to estimate the
unknown nonlogarithmic third-order contribution to the wave function.

In the next section we present the ${\cal O}(\beta_0^3\alpha_s^3)$
corrections for the states with principle quantum number
$n=1,~2,~3$ and angular momentum $l=0$.  In
Section~\ref{sec2} we generalize the complete ${\cal O}(m_q\alpha_s^5)$
result for the ground state energy \cite{PenSte} to the excited
states.  In Section~\ref{sec3} we discuss the impact of the corrections
on the phenomenology of the  $b\bar b$ and $t\bar t$ systems.
Our summary is presented in Section~\ref{sec4}.


\section{Heavy quarkonium parameters to ${\cal O}(\beta_0^3\alpha_s^3)$}
\label{sec1}

In the framework of nonrelativistic effective theory
\cite{CasLep,BBL,PinSot1,BenSmi} the corrections to the heavy quarkonium
parameters are obtained by evaluating the corrections to the Green
function of the effective Schr\"odinger equation \cite{KPSS1}.
The
$\beta_0^3$ part of the third-order contribution results from the
leading renormalization group running of the static potential
which enters the corresponding effective Hamiltonian and is given by
(see also Ref.~\cite{Brambilla:1999qa})
\begin{eqnarray}
  V_C(r)&=& -\frac{C_F\alpha_s}{r}\Bigg\{
  1 
  + \frac{\alpha_s}{4\pi}
  \left(8\beta_0\LrmuEuler + a_1  \right) 
  + \left(\frac{\alpha_s}{4\pi}\right)^2
  \Bigg[ 64\beta_0^2\LrmuEuler^2
    +\left(16 a_1 \beta_0 + 32\beta_1\right)\LrmuEuler 
  \nonumber\\&&
     + a_2 +  \frac{16\pi^2}{3}\beta_0^2
    \Bigg]
  + \left(\frac{\alpha_s}{4\pi}\right)^3
  \Bigg[512\beta_0^3\LrmuEuler^3+\left(192 a_1\beta_0^2
    + 640\beta_0\beta_1\right)\LrmuEuler^2
    \nonumber\\&&
     +\left( 128\pi^2\beta_0^3+24 a_2\beta_0 + 64 a_1 \beta_1 
    + 128\beta_2 + 16\pi^2C_A^3\right) \LrmuEuler 
    \nonumber\\&& 
    + a_3 + 16\pi^2 a_1 \beta_0^2 
    + 1024\zeta(3)\beta_0^3 + \frac{160\pi^2}{3}\beta_0\beta_1 
   \Bigg] 
  + {\cal O}(\alpha_s^4)
  \Bigg\}
  \,,
  \label{eq::Vc}
\end{eqnarray}
where $\LrmuEuler = \ln(e^{\gamma_E}\mu r)$, $\gamma_E=0.577216\ldots$
is Euler's constant, $\zeta(z)$ is Riemann's zeta-function with value
$\zeta(3)=1.202057\ldots$, $C_F=(N_c^2-1)/(2N_c)$ and $C_A=N_c$ for the
$SU(N_c)$ gauge group.  Furthermore, we have
$\alpha_s\equiv\alpha_s(\mu)$ if not stated otherwise.  The coefficients
$a_i$ ($i=1,2$) and $\beta_i$ ($i=0,1,2$) are given in
Appendix~\ref{app::coef}. For the three-loop coefficient $a_3$ only 
Pad\'e estimates are available so far \cite{ChiEli}.  In the order of
interest one has to consider single iterations of the $\beta_0^3$ term,
double iterations of the $\beta_0^2$ and $\beta_0$ term and triple
iterations of the first-order corrections proportional to $\beta_0$.
For the practical computation we use the method elaborated in
Refs.~\cite{KPP,PenPiv1,PenPiv2}.  In this way we obtain the corrections
to the energy levels and wave function at the origin in the form of
multiple harmonic sums.  For general $n$ the result is rather
cumbersome.  For a specific $n$, however, the summation can be performed
analytically.  Below we present our result for $n=1,~2,~3$ and $l=0$
which is sufficient for the phenomenological applications. For vanishing
angular momentum we can write the perturbative part of the energy level
with principal quantum number $n$ as
\begin{eqnarray}
  E_n^{\rm p.t.}&=&E^C_n+\delta E^{(1)}_n+\delta E^{(2)}_n+\delta
  E^{(3)}_n
  +\ldots\,,
  \label{enseries}
\end{eqnarray}
where $\delta E^{(k)}_n$ stands for corrections of order $\alpha_s^k$.
The leading order Coulomb energy is given by
\begin{eqnarray}
  E^C_n&=&-{C_F^2\alpha_s^2m_q\over4n^2}\,.
\end{eqnarray}
For the ${\cal O}(\beta_0^3\alpha_s^3)$ term we obtain
\begin{eqnarray}
  \delta^{(3)}_{\beta_0^3}  E_1&=&
  E^C_1\left({\beta_0\alpha_s\over \pi}\right)^3
  \Bigg[32 \lmu{1}^3 + 40 \lmu{1}^2 + \left({16\pi^2\over 3}+64\zeta(3)
    \right)\lmu{1}
    \nonumber\\&&
    -8+{4\pi^2}+{2\pi^4\over 45}
    +64\zeta(3)-{8\pi^2\zeta(3)}+96\zeta(5)\Bigg]\,, 
  \nonumber\\
  \delta^{(3)}_{\beta_0^3}  E_2
  &=&E^C_2\left({\beta_0\alpha_s\over \pi}\right)^3
  \Bigg[32 \lmu{2}^3+{88}\lmu{2}^2+\left({32}+{16\pi^2\over 3}
    +128\zeta(3)\right)\lmu{2}
    \nonumber\\&&
    -{102}+{52\pi^2\over 3}
    +{4\pi^4\over 45}+112\zeta(3)-{32\pi^2\zeta(3)}+384\zeta(5)\Bigg]\,, 
  \nonumber\\
  \delta^{(3)}_{\beta_0^3}  E_3&=&
  E^C_3\left({\beta_0\alpha_s\over \pi}\right)^3
  \Bigg[32 \lmu{3}^3+{120}\lmu{3}^2
    +\left({136\over 3}+{16\pi^2\over 3}
    +192\zeta(3)\right)\lmu{3}
    \nonumber\\
    &&
    -{9514\over 27}    
    +{427\pi^2\over 9}+{2\pi^4\over 15}+{140\zeta(3)}-{72\pi^2\zeta(3)}
    +864\zeta(5)\Bigg]\,,
\label{Ebt3}
\end{eqnarray}
where $\lmu{n}= \ln(n\mu/(C_F\alpha_s(\mu)m_q))$ 
and $\zeta(5)=1.036927\ldots$.
Note that the $n=1$ result has already been known
\cite{KiySum,Hoa1,PenSte}. The perturbative expansion for the wave
function can be written as follows
\begin{eqnarray}
  |\psi_n(0)|^2&=&|\psi^C_n(0)|^2\left(1+\delta^{(1)}\psi_n
  +\delta^{(2)}\psi_n+\delta^{(3)}\psi_n+\ldots\right)\,,
\end{eqnarray}
where
\begin{eqnarray}
  |\psi^C_n(0)|^2&=&{C_F^3\alpha_s^3 m_q^3\over 8\pi n^3}
  \,,
\end{eqnarray}
is  the leading order Coulomb value. Our result for the
${\cal O}(\beta_0^3\alpha_s^3)$ term reads
\begin{eqnarray}
  \delta^{(3)}_{\beta_0^3} \psi_1&=&\left({\beta_0\alpha_s\over
    \pi}\right)^3
  \left[80\lmu{1}^3+\left({52}-{80\pi^2\over 3}\right)\lmu{1}^2
    +\left(-40-{6\pi^2}+{10\pi^4\over 9}
    +200\zeta(3)\right)\lmu{1}\right.
    \nonumber\\
    &&
    -{20}+{22\pi^2\over 3}
    -{7\pi^4\over 5}+{4\pi^6\over 105}
    +112\zeta(3)-{12\pi^2\zeta(3)}-16\zeta(3)^2-40\zeta(5)\Bigg]
    \,, 
    \nonumber\\
    \delta^{(3)}_{\beta_0^3} \psi_2&=&\left({\beta_0\alpha_s\over
      \pi}\right)^3 
    \left[80\lmu{2}^3+\left({332}-{160\pi^2\over 3}\right)\lmu{2}^2
      +\left({308}-{266\pi^2\over 3}+{40\pi^4\over 9}
      +400\zeta(3)\right)\lmu{2}\right.
      \nonumber\\
      &&
      -{361}+{73\pi^2\over 3}
      -{26\pi^4\over 45}+{32\pi^6\over 105}
      +496\zeta(3)-{48\pi^2\zeta(3)}-128\zeta(3)^2-160\zeta(5)\Bigg]
      \,, 
      \nonumber\\
      \delta^{(3)}_{\beta_0^3} \psi_3&=&
      \left({\beta_0\alpha_s\over \pi}\right)^3 
      \left[80\lmu{3}^3+\left({612}-80\pi^2\right)\lmu{3}^2
        +\left({2893\over 3}-{228\pi^2}+{10\pi^4}
        +600\zeta(3)\right)\lmu{3}\right.
        \nonumber\\
        &&
        -{100679\over 54}+{183\pi^2\over 2}
        +{52\pi^4\over 15}+{36\pi^6\over 35}
        +{1374\zeta(3)}
	-{108\pi^2\zeta(3)}
	-432\zeta(3)^2
        \nonumber\\
        &&-360\zeta(5)\Bigg]
        \,.
\end{eqnarray}


\section{Exited states spectrum to ${\cal O}(m_q\alpha_s^5)$}
\label{sec2}
The heavy quarkonium spectrum up to ${\cal O}(m_q\alpha_s^4)$ has been
derived in Refs.~\cite{PinYnd,PenPiv1,MelYel1}.  For convenience of
the reader the
expressions for $\delta E_n^{(1)}$ and $\delta E_n^{(2)}$ are listed in
Appendix~\ref{app::results}.  
At ${\cal O}(m_q\alpha_s^5)$ it is convenient to split
$\delta E_n^{(3)}$ into two parts: one corresponding to vanishing
beta-function and one proportional to the coefficients of
beta-function:
\begin{eqnarray}
  \delta E_n^{(3)} &=& 
  \delta E_n^{(3)}\Big|_{\beta(\alpha_s)=0}
  + \delta E_n^{(3)}\Big|_{\beta(\alpha_s)} 
  \,.
\end{eqnarray}
The contribution $\delta E_n^{(3)}\Big|_{\beta(\alpha_s)=0}$ has been
evaluated in Ref.~\cite{KPSS1}. For completeness we
include the corresponding expressions in Appendix B.  In Ref.~\cite{PenSte}
the quantity $\delta E_n^{(3)}\Big|_{\beta(\alpha_s)}$ has been
computed for $n=1$. Below we extend it to the excited states.  Following
Ref.~\cite{PenSte} we divide $\delta E_n^{(3)}\Big|_{\beta(\alpha_s)}$
into four pieces
\begin{eqnarray}
  \delta E_n^{(3)}\Big|_{\beta(\alpha_s)} &=& 
  \delta E_n^{(3)}\Big|_{\rm C.r.} 
  + \delta E_n^{(3)}\Big|_{\rm B.r.} 
  + \delta E_n^{(3)}\Big|_{\rm C.i.} 
  + \delta E_n^{(3)}\Big|_{\rm B.i.} 
  \,.
  \label{eq::En3beta}
\end{eqnarray}
The first two terms of the above equation are related to the running of
the lower-order potentials.  The contribution $\delta
E_n^{(3)}\Big|_{\rm C.r.}$ is due to the three-loop running of the
static potential, Eq.~(\ref{eq::Vc}). It reads
\begin{eqnarray}
  \delta E_n^{(3)}\Big|_{\rm C.r.}  &=&
  E_n^C \left(\frac{\alpha_s(\mu)}{\pi}\right)^3 \Bigg\{
  \left(6 a_1\beta_0^2 + 20\beta_0\beta_1\right)\lmu{n}^2 
  + \left(12 \PolyEuler{n+1}a_1\beta_0^2+\frac{3}{4}a_2 \beta_0 + 2 a_1\beta_1 
  \right.\nonumber\\&&\mbox{}
  + 40\PolyEuler{n+1} \beta_0\beta_1 + 4\beta_2 
  \bigg)\lmu{n}
 +
  \left(-\frac{12}{n^2} + \frac{5\pi^2}{2} 
  - \frac{12}{n}\PolyEuler{n} + 6\PolyEuler{n+1}^2 
  - 6\Psi_2(n+1)
  \right)
  \nonumber\\&&\mbox{}
  \times a_1\beta_0^2+\frac{3}{4}\PolyEuler{n+1}a_2\beta_0
  +2\PolyEuler{n+1}a_1\beta_1  + \left(-\frac{40}{n^2} + \frac{25\pi^2}{3} 
  - \frac{40}{n}\PolyEuler{n} + 20\PolyEuler{n+1}^2 
  \right.
   \nonumber\\&&\mbox{}
  - 20\Psi_2(n+1)\Bigg)\beta_0\beta_1
  + 4\PolyEuler{n+1}\beta_2 \Bigg\}
  + \delta^{(3)}_{\beta_0^3} E_{n}\Big|_{\rm C.r.}
  \,,
  \label{eq::En3Cr}
\end{eqnarray}
where $\PolyEuler{n}=\Psi_1(n)+\gamma_E$, $\Psi_n(z)={\rm
  d}^n\ln(\Gamma(z))/{\rm d} z^n$ and $\Gamma(z)$ is the Euler's
gamma-function.  The term
$\delta^{(3)}_{\beta_0^3} E_{n}\Big|_{\rm C.r.}$
in Eq.~(\ref{eq::En3Cr}) is included in Eq.~(\ref{Ebt3}).  

The contribution $\delta
E_n^{(3)}\Big|_{\rm B.r.}$ is due to the one-loop running of the power
suppressed terms in the NNLO\footnote{LO, NLO, $\ldots$ stand for the
  leading order, next-to-leading order, {\it etc}.} effective
Hamiltonian (see, e.g., Ref.~\cite{KPSS1}), which we denote as the
``Breit potential''.  For this contribution we obtain
\begin{eqnarray}
  \delta E_n^{(3)}\Big|_{\rm B.r.}  &=&
  E_n^C \frac{\alpha_s^3(\mu)}{\pi} \beta_0 \Bigg\{
  \Bigg[ \frac{4}{n} C_FC_A
   +\left(2-\frac{1}{n} -\frac{4}{3}\sss\right)\frac{C_F^2}{n} 
    \Bigg] \lmu{n}
  \nonumber\\&&\mbox{}
  +\left( 4 - 4\PolyEuler{n+1} \right)  \frac{C_FC_A}{n} 
  + \left[\left(2 -\left(2+\frac{1}{n}\right) \PolyEuler{n+1}
  \right)\right.
  \nonumber\\&&\mbox{}\left.
  + \left(
  -\frac{2}{3n} + \frac{2}{3} + \frac{4}{3}\PolyEuler{n+1}
  \right)\sss \right]  \frac{C_F^2}{n}
    \Bigg\}
  \,,
  \label{eq::En3Br}
\end{eqnarray}
where $S$ is the spin quantum number.

The remaining two contributions of Eq.~(\ref{eq::En3beta}) are related
to the iteration of lower-order potentials.  The contribution $\delta
E_n^{(3)}\Big|_{\rm C.r.}$ corresponds to the iteration of the one- and
two-loop running of the static potential and is of the following form
\begin{eqnarray}
  \delta E_n^{(3)}\Big|_{\rm C.i.} \!\! &=&\!\!
  E_n^C \left(\frac{\alpha_s(\mu)}{\pi}\right)^3 \Bigg\{
  \left(6a_1\beta_0^2 + 8\beta_0\beta_1 \right) \lmu{n}^2 
  + \Bigg[\frac{a_1^2\beta_0}{2} + \frac{a_2\beta_0}{4}
  \nonumber\\&&\mbox{}
  +\left(-14 + \frac{12}{n} + 12\PolyEuler{n}\right)  
  a_1\beta_0^2+a_1\beta_1   
  +\left(-16 + \frac{16}{n} + 16\PolyEuler{n}\right)
  \beta_0\beta_1
  \Bigg]\lmu{n} \nonumber\\&&\mbox{}
    + \left(-\frac{5}{8} + \frac{1}{2n} 
    + \frac{\PolyEuler{n}}{2}\right) a_1^2 \beta_0+\left(
    -\frac{1}{4} + \frac{1}{4n}+ \frac{\PolyEuler{n}}{4}\right) a_2 \beta_0
  +\left[2 + \frac{12}{n^2} - \frac{14}{n}  + \frac{5\pi^2}{6}\right.
 \nonumber\\&&\mbox{}\left.
  + \left(-14+\frac{16}{n}\right)\PolyEuler{n} + 6\PolyEuler{n}^2 
  - 10\Psi_2(n)  - 4n \Psi_3(n)
  \right] a_1\beta_0^2
  +\left(-1 + \frac{1}{n} + \PolyEuler{n}\right)  
  \nonumber\\&&\mbox{}
   \times a_1\beta_1  +  \left[\frac{24}{n^2} 
    - \frac{16}{n} 
    + \left(-16 + \frac{32}{n}\right)\PolyEuler{n}
    + 8\PolyEuler{n}^2 
    - 16\Psi_2(n) - 8 n\Psi_3(n)
    \right]\beta_0 \beta_1
  \Bigg\}\nonumber\\&&\mbox{}
  + \delta^{(3)}_{\beta_0^3} E_{n}\Big|_{\rm C.i.}
  \,,
  \label{eq::En3Ci}
\end{eqnarray}
where $\delta^{(3)}_{\beta_0^3} E_{n}\Big|_{\rm C.i.}$
contributes to Eq.~(\ref{Ebt3}).  

The last contribution $\delta E_n^{(3)}\Big|_{\rm
  B.i.}$ incorporates the iteration of the Breit potential and the
one-loop running of the static potential.  It reads
\begin{eqnarray}
  \delta E_n^{(3)}\Big|_{\rm B.i.}  &=&
  E_n^C \frac{\alpha_s^3(\mu)}{\pi} \beta_0 \Bigg\{
  \Bigg[\frac{4}{n}  C_FC_A 
    +\left(-\frac{9}{2n} + 14 - 4\sss \right)\frac{C_F^2}{n}
    \Bigg]\lmu{n}
  \nonumber\\&&\mbox{}
  +\left(\frac{4}{n^2} - \frac{2}{n} 
  + \frac{4}{n}\PolyEuler{n+1} - 4\Psi_2(n)
  \right) C_FC_A 
  + \left[\frac{19}{2n^2} + \frac{2}{n} 
  +\left(-\frac{9}{2n^2} + \frac{2}{n}\right)  \PolyEuler{n+1}
  \right.
  \nonumber\\&&\mbox{}\left.
   - 8\Psi_2(n)+\left(-\frac{8}{3n^2} + \frac{4}{3n} 
  - \frac{4}{3n}\PolyEuler{n+1} + \frac{8}{3}\Psi_2(n)
  \right)\sss\right] C_F^2
  \Bigg\}
  \,.
  \label{eq::En3Bi}
\end{eqnarray}
After summing up the four contributions according to
Eq.~(\ref{eq::En3beta}) we obtain our final result for the ${\cal
  O}(\alpha_s^3)$ corrections to the energy levels 
involving coefficients of the beta function.
For $n=1,~2,$ and 3 they read
\begin{eqnarray}
  \lefteqn{\delta E_1^{(3)}\Big|_{\beta(\alpha_s)} = 
  E_1^C \left(\frac{\alpha_s(\mu)}{\pi}\right)^3 \Bigg\{
  32\beta_0^3\lmu{1}^3
  + \left( 12a_1\beta_0^2 + 40\beta_0^3 + 28\beta_0\beta_1 \right)\lmu{1}^2 
  }
  \nonumber\\&&\mbox{}
  + \Bigg[\frac{a_1^2 \beta_0}{2} + a_2 \beta_0
  + 10a_1\beta_0^2+ \left(\frac{16\pi^2}{3} + 64\zeta(3)\right)\beta_0^3
  + 3a_1\beta_1 + 40\beta_0\beta_1  + 4\beta_2 
 \nonumber\\&&\mbox{} + 8\pi^2\beta_0C_FC_A
  + \left(\frac{21}{2}
  - \frac{16}{3}\sss
  \right) \pi^2\beta_0C_F^2
  \Bigg] \lmu{1} -\frac{a_1^2\beta_0}{8} + \frac{3}{4}a_2\beta_0
  +\left(\frac{2\pi^2}{3} + 8\zeta(3)\right)
  \nonumber\\&&\mbox{} \times a_1\beta_0^2
  +\left[-8 + \frac{2\pi^4}{45} 
  +\left(4 - 8\zeta(3)\right) \pi^2 + 64\zeta(3) + 96\zeta(5)\right]\beta_0^3
  + 2 a_1\beta_1
 \nonumber\\&&\mbox{}+ \left(8
  + \frac{7\pi^2}{3} + 16\zeta(3)\right)\beta_0\beta_1 + 4\beta_2 
  + \left(6 - \frac{2\pi^2}{3}\right) \pi^2  \beta_0C_FC_A
  + \left[8 - \frac{4\pi^2}{3} \right.
  \nonumber\\&&\mbox{}\left.
  + \left(-\frac{4}{3} + \frac{4\pi^2}{9}\right)\sss\right]\pi^2\beta_0C_F^2
  \Bigg\}
  \,,
  \nonumber\\
\lefteqn{\delta E_2^{(3)}\Big|_{\beta(\alpha_s)} = 
  E_2^C \left(\frac{\alpha_s(\mu)}{\pi}\right)^3 \Bigg\{
  32\beta_0^3\lmu{2}^3
  + \left( 12a_1\beta_0^2 + 88\beta_0^3 + 28\beta_0\beta_1 \right)\lmu{2}^2 
  }
  \nonumber\\&&\mbox{}
  + \Bigg[\frac{a_1^2 \beta_0}{2} + a_2 \beta_0
  + 22a_1\beta_0^2+ \left(32+\frac{16\pi^2}{3} + 128\zeta(3)\right)\beta_0^3
  + 3a_1\beta_1 + 68\beta_0\beta_1  + 4\beta_2 
 \nonumber\\&&\mbox{} + 4\pi^2\beta_0C_FC_A
  + \left(\frac{53}{8}
  - \frac{8}{3}\sss
  \right) \pi^2\beta_0C_F^2
  \Bigg] \lmu{2} +\frac{a_1^2\beta_0}{8} + \frac{5}{4}a_2\beta_0
  +\left(4+\frac{2\pi^2}{3}\right.
  \nonumber\\&&\mbox{}  + 16\zeta(3)\Bigg)a_1\beta_0^2
  +\left[-102 + \frac{4\pi^4}{45} 
  + \left(\frac{52}{3} - 32\zeta(3)\right)\pi^2 + 112\zeta(3) 
  + 384\zeta(5)\right]\beta_0^3
 \nonumber\\&&\mbox{}
  + \frac{7}{2} a_1\beta_1
 + \left(30
  + \frac{7\pi^2}{3} + 32\zeta(3)\right)\beta_0\beta_1 + 6\beta_2 
  + \left(6 - \frac{2\pi^2}{3}\right) \pi^2  \beta_0C_FC_A
  + \left[\frac{165}{16} - \frac{4\pi^2}{3} \right.
  \nonumber\\&&\mbox{}\left.
  + \left(-\frac{5}{2} + \frac{4\pi^2}{9}\right)\sss\right]\pi^2\beta_0C_F^2
  \Bigg\}
  \,,
  \nonumber\\
\lefteqn{\delta E_3^{(3)}\Big|_{\beta(\alpha_s)} = 
  E_3^C \left(\frac{\alpha_s(\mu)}{\pi}\right)^3 \Bigg\{
  32\beta_0^3\lmu{3}^3
  + \left( 12a_1\beta_0^2 + 120\beta_0^3 + 28\beta_0\beta_1 \right)\lmu{3}^2 
  }
  \nonumber\\&&\mbox{}
  + \Bigg[\frac{a_1^2 \beta_0}{2} + a_2 \beta_0
  + 30a_1\beta_0^2+ \left(\frac{136}{3}
  +\frac{16\pi^2}{3} + 192\zeta(3)\right)\beta_0^3
  + 3a_1\beta_1 + \frac{260}{3}\beta_0\beta_1  + 4\beta_2 
 \nonumber\\&&\mbox{} + \frac{8\pi^2}{3}\beta_0C_FC_A
  + \left(\frac{85}{18}
  - \frac{16}{9}\sss
  \right) \pi^2\beta_0C_F^2
  \Bigg] \lmu{3} +\frac{7}{24}a_1^2\beta_0 + \frac{19}{12}a_2\beta_0
  +\left(\frac{17}{3} +\frac{2\pi^2}{3} \right.
  \nonumber\\&&\mbox{} + 24\zeta(3)\Bigg) a_1\beta_0^2
  +\left[- \frac{9514}{27} + \frac{2\pi^4}{15} 
  +\left(\frac{427}{9} - 72\zeta(3)\right) \pi^2 + 140\zeta(3) 
  + 864\zeta(5)\right]\beta_0^3
 \nonumber\\&&\mbox{} +  \frac{9}{2} a_1\beta_1+ \left( \frac{130}{3}
  + \frac{7\pi^2}{3} + 48\zeta(3)\right)\beta_0\beta_1 +  \frac{22}{3}\beta_2 
  + \left( \frac{55}{9} - \frac{2\pi^2}{3}\right) \pi^2  \beta_0C_FC_A
  \nonumber\\&&\mbox{} + \left[ \frac{1217}{108} - \frac{4\pi^2}{3}
  + \left(-\frac{82}{27} + \frac{4\pi^2}{9}\right)\sss\right]\pi^2\beta_0C_F^2
  \Bigg\}
  \,.
  \label{eq::En3beta_fin}
\end{eqnarray}
The equation with $n=1$ agrees with the result of Ref.~\cite{PenSte}.
The Eqs.~(\ref{eq::En3beta_fin}),~(\ref{eq::En1}),~(\ref{eq::En2}) 
and~(\ref{spectrum}) provide 
the complete result for the energy levels up to ${\cal O}(m_q\alpha_s^5)$. 
We should note that, although we only present analytical results for the
first three principle quantum numbers, there is no principle 
problem to
obtain expressions for higher excited states, too.  However, from the
phenomenological point of view they are far less important, and thus
we refrain from listing them explicitly.  

It is
instructive to evaluate the energy levels in numerical form:
\begin{eqnarray}
  \frac{\delta E^{(3)}_1}{E_1^C} \!\!\!&=&\!\!\! \alpha_s^3\left[
    \left(
    \begin{array}{c} 
      70.590|_{n_l=4}\\
      56.732|_{n_l=5} 
    \end{array} 
    \right)
    + 15.297 \ln\alpha_s 
    + 0.001\,a_3 +  \left.\left(
    \begin{array}{c} 
      34.229|_{n_l=4}\\
      26.654|_{n_l=5} 
    \end{array} 
    \right)\right|_{\beta_0^3}
    \right]\,,
  \nonumber
  \\
  \frac{\delta E^{(3)}_2}{E_2^C} \!\!\!&=&\!\!\! \alpha_s^3\left[
    \left(
    \begin{array}{c} 
      84.634|_{n_l=4}\\
      62.164|_{n_l=5} 
    \end{array} 
    \right)
    + 8.647 \ln\alpha_s 
    + 0.001\,a_3 +  \left.\left(
    \begin{array}{c} 
      67.337|_{n_l=4}\\
      52.434|_{n_l=5} 
    \end{array} 
    \right)\right|_{\beta_0^3}
    \right]\,,
  \nonumber
  \\
  \frac{\delta E^{(3)}_3}{E_3^C} \!\!\!&=&\!\!\! \alpha_s^3\left[
    \left(
    \begin{array}{c} 
      101.69|_{n_l=4}\\
      72.368|_{n_l=5} 
    \end{array} 
    \right)
    +  6.305\ln\alpha_s 
    + 0.001\,a_3 +  \left.\left(
    \begin{array}{c} 
      98.824|_{n_l=4}\\
      76.953|_{n_l=5} 
    \end{array} 
    \right)\right|_{\beta_0^3}
    \right]\,,
\label{Enum}
\end{eqnarray}
where $\alpha_s=\alpha_s(\mu_s/n)$ and $\mu=\mu_s/n$ 
with $\mu_s=C_F \alpha_s(\mu_s) m_q$
and we put $S=1$ which corresponds to
the spin-triplet state.  The recent analysis of the spin-dependent
contribution to the spectrum, which is 
responsible for the hyperfine splitting, can
be found in Refs.~\cite{KPPSS,PPSS}. In Eq.~(\ref{Enum}) we have
separated the contributions arising from $a_3$ and $\beta_0^3$.  Using
the Pad\'e estimates~\cite{ChiEli} we obtain $0.001\,a_3|_{n_l=4}\approx
6$ and $0.001\,a_3|_{n_l=5}\approx 4$.  Thus, the result for the energy
levels depends only marginally on the precise value of $a_3$ provided
the Pad\'e estimates give the correct order of magnitude.
Furthermore, one can see that the $\beta_0^3$ term contributes between
25\% ($n=1$) and 50\% ($n=3$) of the nonlogarithmic term.


\section{Heavy quarkonium phenomenology}
\label{sec3}

In this section we discuss some phenomenological applications of the
results derived in the previous parts of the paper. As input values for the
numerical analyses we adopt $\alpha_s(M_Z)=0.118$, and
$m_b=5.3$~GeV and $m_t=175$~GeV for the quark pole masses. 
Furthermore, we use the soft scale  $\mu_s \approx 2.10$~GeV 
for the bottom and $\mu_s \approx 32.6$~GeV for the top
quark case.


\vspace*{1em}
\noindent
{\bf Excited states of bottomonium.}
The mass of the $\Upsilon(nS)$ meson can be decomposed into perturbative
and nonperturbative contributions
\begin{eqnarray}
  M_{\Upsilon(nS)} &=&
  2m_b + E_n^{\rm p.t.}+ \delta^{\rm n.p.}E_n\,.
  \label{mqq}
\end{eqnarray}
The perturbative contribution $E_n^{\rm p.t.}$ up to ${\cal O}(m_q\alpha_s^5)$
is given in the previous sections. The phenomenological application of the 
result to the $\Upsilon(1S)$ meson mass 
has been discussed in Ref.~\cite{PenSte}. For the exited states
let us consider the ratio
\begin{eqnarray}
  \rho_n&=&{E_n-E_1\over 2m_b+E_1}\,.
  \label{rho}
\end{eqnarray}
It depends on the quark mass only through the normalization
scale of $\alpha_s$ and does not suffer from renormalon contributions.
Including successively higher orders  one gets for $\mu=\mu_s$
\begin{eqnarray}
 10^2 \times\rho^{\rm p.t.}_2 &=& 1.49\left(
  1 +0.79_{\rm NLO} + 1.18_{\rm NNLO} + 1.21_{\rm N^3LO}+\ldots
  \right)
  \,,
  \nonumber\\
 10^2 \times\rho^{\rm p.t.}_3  &=& 1.77\left(
  1 + 0.92_{\rm NLO} + 1.37_{\rm NNLO} + 1.55_{\rm N^3LO}+\ldots
  \right)
  \,,
  \label{eq::rho}
\end{eqnarray}
where $\alpha_s^{(4)}(\mu_s)$ is extracted from its value at $M_Z$ using
four-loop beta-function accompanied with three-loop
matching\footnote{We use the package {\tt
RunDec}~\cite{Chetyrkin:2000yt} to perform the running and
matching of $\alpha_s$.}.
Though the convergence of the series is not good, the N$^3$LO
perturbative result is in impressive agreement with the experimental
values $\rho^{\rm exp}_n =(M_{\Upsilon(nS)}-M_{\Upsilon(1S)})/
M_{\Upsilon(1S)}$ for $n=2$ and $3$ as can be seen in Tab.~\ref{tab1}.
We would like to
emphasize the role of the perturbative corrections necessary to
bring theory and experiment into agreement which we will use in the
following to estimate the order of magnitude
of the nonperturbative effects. In fact the absence of a sufficiently
accurate estimate of the nonperturbative part $\delta^{\rm n.p.}E_n$ is
one of the main problems in the theory of heavy quarkonium.  In the
limit $\alpha_s^2m_q\gg \Lambda_{QCD}$ it can be investigated by the
method of vacuum condensate expansion \cite{Vol,Leu}. However, for
bottomonium it can only be used for $n=1$.  For higher states the
leading term due to the gluonic condensate grows as $n^6$. It becomes
unacceptably large already for $n=2$ where the whole series blows up
\cite{Pin1}. Even for $n=1$ such an estimate suffers from large
uncertainties due to the poorly known value of the gluonic condensate 
and due to a strong
scale dependence. A rough numerical estimate is $\delta^{\rm
  n.p.}E_1\approx 60$~MeV~\cite{PenSte}. Since our perturbative 
result agrees very well with the experimental result
we can conclude that $\delta^{\rm n.p.}E_2$ should be of the same
size as $\delta^{\rm n.p.}E_1$.
In general for
bottomonium the nonperturbative corrections
appear to be rather moderate and the
theoretical estimates are dominated by perturbative contributions.
Similar conclusion has been made in Ref.~\cite{BVS} in a somewhat
different framework.

\begin{table}[t]
\begin{center}
\renewcommand{\arraystretch}{1.2}
\begin{tabular}{|l||c|c|}
  \hline
  & $\Upsilon(2S)$ & $\Upsilon(3S)$ \\
  \hline
  $ 10^2 \times\rho^{\rm p.t.}_n$ 
  & $6.2^{+1.7}_{-1.2}$ & $8.6^{+2.4}_{-1.8}$ \\
  $ 10^2 \times\rho^{\rm exp}_n$  
  & $5.95$& $9.46$\\
  \hline
\end{tabular}
\caption{\label{tab1}
  Perturbative versus experimental results for the parameter $\rho_n$
  as defined in Eq.~(\ref{rho}).
  The theoretical uncertainty corresponds to
  $\alpha_s(M_Z)=0.118\pm 0.003$. The experimental values are extracted from
  Ref.~\cite{Hag}.
  For $a_3$ we used the Pad\'e
  estimate~\cite{ChiEli}  $a_3\big|_{n_l=4}=6272$.}
\end{center}
\end{table}

\begin{figure}[t]
  \begin{tabular}{cc}
    \epsfxsize=8.5cm 
    \epsfig{figure=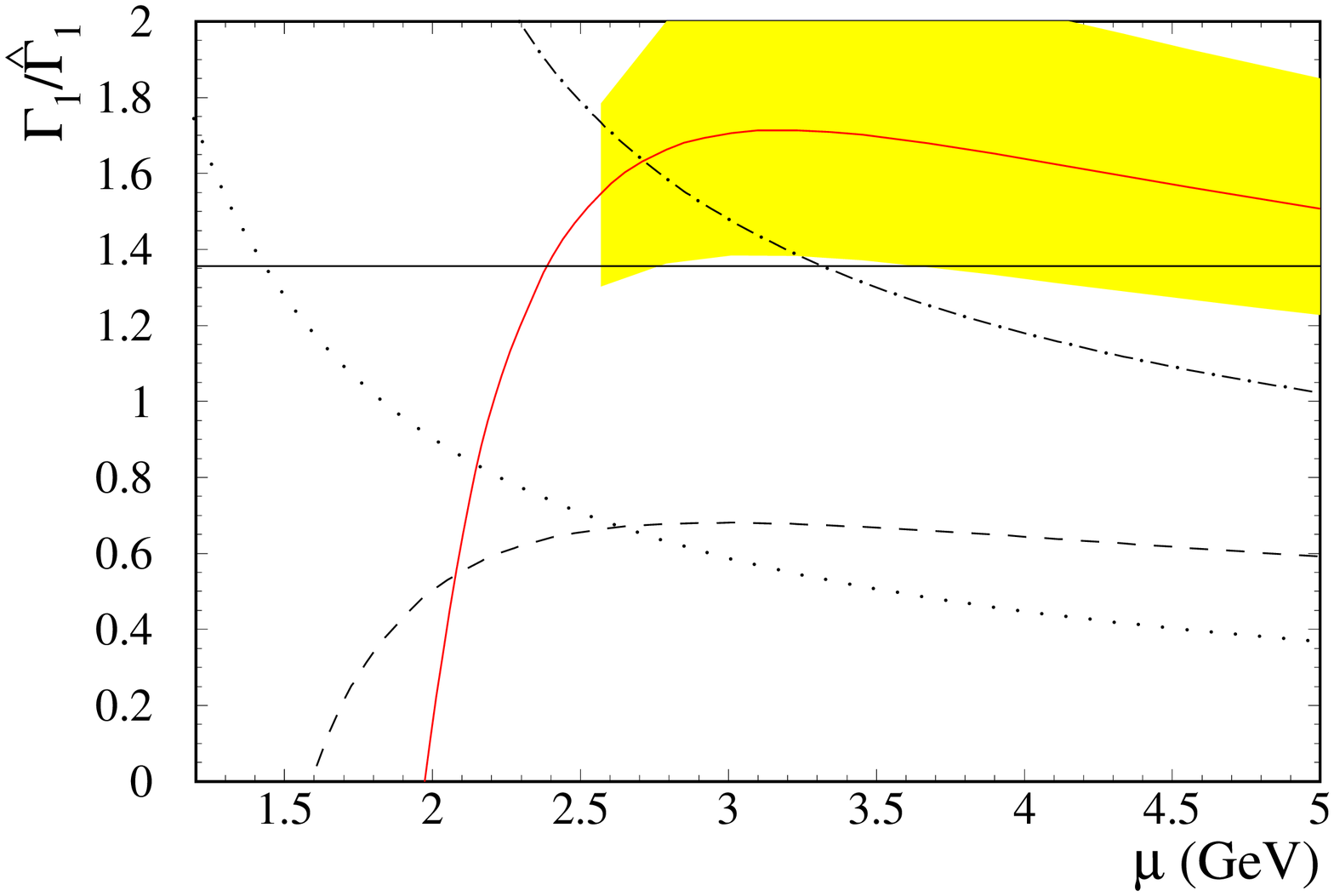,width=7.5cm}
    &
    \epsfxsize=8.5cm 
    \epsfig{figure=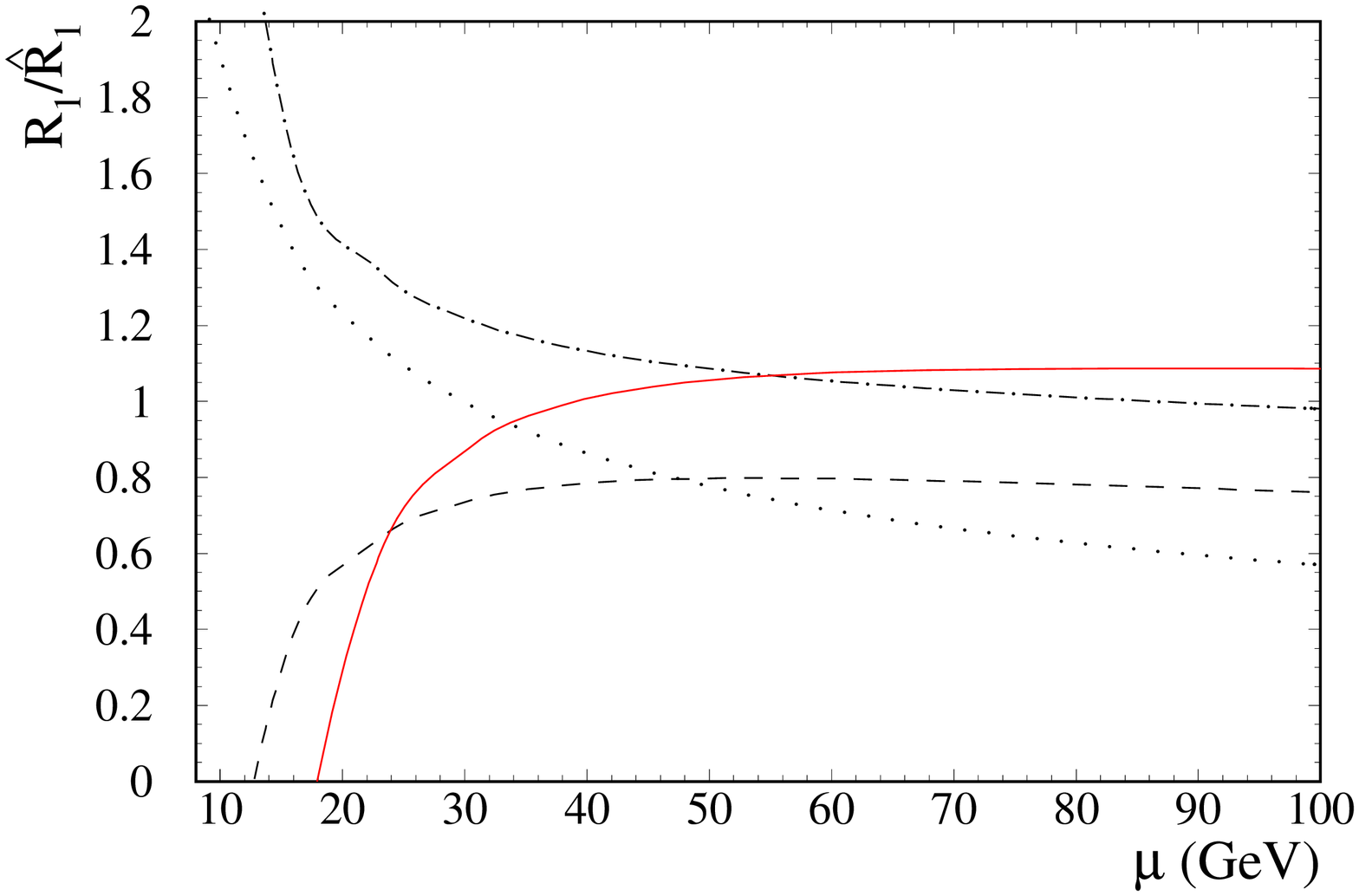,width=7.5cm}
    \\
    (a) & (b)
  \end{tabular}
  \caption{\label{fig::mu} (a) $\Gamma_1$ normalized to 
    $\hat{\Gamma}_1\equiv\Gamma_1^{\rm LO}\big|_{\alpha_s\to \alpha_s(\mu_s)}$ 
    as a function of $\mu$ at LO (dotted), NLO (dashed), NNLO
    (dotted-dashed) and N$^3$LO$'$ (full line).  The horizontal line
    corresponds to the experimental value $\Gamma^{\rm
      exp}(\Upsilon(1S)\to e^+e^-)=1.31$~keV \cite{Hag}. 
    For the N$^3$LO$'$ result, the band
    reflects the errors due to $\alpha_s(M_Z)=0.118\pm 0.003$.  
    (b) The analog plot for $R_1$ with 
    $\hat{R}_1\equiv R_1^{\rm LO}\big|_{\alpha_s\to \alpha_s(\mu_s)}$.}
\end{figure}

\vspace*{1em}
\noindent
{\bf $\Upsilon(1S)$ leptonic width.}
In the nonrelativistic effective theory the leading order approximation
for the leptonic width $\Gamma^{\rm LO}(\Upsilon(1S)\to l^+l^-)\equiv
\Gamma_1$ reads $\Gamma^{\rm LO}_1=4\pi N_cQ_b^2\alpha^2|\psi_1^C(0)|^2/$
$\left(3m_b^2\right)$, with $N_c=3$ and $Q_b=-1/3$. Combining the known
perturbative results up to ${\cal O}(\alpha_s^3\ln\alpha_s)$ (see
Ref.~\cite{KPSS2}) with the ${\cal O}(\beta_0^3\alpha_s^3)$ contribution
obtained in Section~\ref{sec1} we obtain the following series
\begin{eqnarray}
  \Gamma_1 
  &\approx& \Gamma_1^{\rm LO}
  \left(1
  -1.70\,\alpha_s(m_b)
  -7.98\,\alpha_s^2(m_b)+\ldots\right)
  \nonumber\\
  &&{}\times\left(1
  -0.30\,\alpha_s
  -5.19\,\alpha_s^2\ln\alpha_s
  +17.2\,\alpha_s^2
  \right. \nonumber\\&&{}\left.
  -14.4\,\alpha_s^3\ln^2\alpha_s
  +0.17\,\alpha_s^3\ln\alpha_s
  -34.9\,\alpha_s^3|_{\beta_0^3}
  +\ldots\right)
  \,,
  \label{eq::gamser}
\end{eqnarray}
where $\alpha_s=\alpha_s(\mu_s)$.
The contribution coming from the hard virtual momenta
region \cite{CzaMel,BSS} is separated and the corresponding strong coupling 
is normalized at
$\mu=m_b$.  Evaluating Eq.~(\ref{eq::gamser}) and retaining only the
logarithmic and $\beta_0^3$ terms at N$^3$LO we find
\begin{eqnarray}
  \Gamma_1 &\approx& \Gamma_1^{\rm LO}(1 -0.445_{\rm NLO}
  +1.75_{\rm NNLO}-1.67_{\rm N^3LO^\prime}+\ldots)
  \,,
  \label{gamnum}
\end{eqnarray}
where the prime indicates that the N$^3$LO corrections are not complete.
Though the perturbative corrections are huge, the rapid growth of the
perturbative coefficients stops at NNLO if we assume that the
$\beta_0^3$ term sets the scale of the nonlogarithmic third-order
contribution.  In Fig.~\ref{fig::mu}(a), the width is plotted as a
function of $\mu$ including the LO, NLO, NNLO and N$^3$LO$'$ approximations
along with the experimental value.  
For the numerical evaluation we extract
$\alpha_s^{(4)}(m_b)$ from its value at $M_Z$ using
four-loop beta-function accompanied with three-loop
matching.
$\alpha_s^{(4)}(m_b)$ is used as starting point in order to evaluate
$\alpha_s^{(4)}(\mu)$ at N$^k$LO
with the help of the $(k+1)$-loop beta-function.
As one can see in Fig.~\ref{fig::mu}(a), the available ${\cal
  O}(\alpha_s^3)$ terms stabilize the series and significantly 
reduce the scale
dependence.  At the scale $\mu'\approx 2.7$~GeV, which is close
to the physically motivated scale $\mu_s$, the N$^3$LO$'$ corrections
vanish and at the scale $\mu''\approx 3.1$~GeV the result becomes
independent of $\mu$; {\it i.e.}, the N$^3$LO$'$ curve shows a local
maximum. In the whole range of $\mu$ between 2~GeV and 5~GeV the result
for the width agrees with the experimental value within the error bar
due to the uncertainty of the strong coupling constant. This may
signal that the missing perturbative corrections are rather moderate.
Furthermore, this result constitutes a significant improvement as
compared to the NLL approximation discussed in
Ref.~\cite{Pineda:2003be}.

For a definite conclusion, however,
one has to wait until the third-order corrections are completed.  The
potentially most important part to be computed is
the ultrasoft contribution which includes $\alpha_s(\mu)$ normalized at
relatively low ultrasoft scale $\mu_{us}\sim \alpha_s^2m_q$.  Currently
only a partial result for this contribution exists \cite{KniPen1}.

\vspace*{1em}
\noindent
{\bf $\Upsilon$ sum rules.}
The nonrelativistic $\Upsilon$ sum rules~\cite{NSVZ} 
operate with the high moments  of the spectral density with  
$n\sim 1/\alpha_s^2$, which are saturated by the nonrelativistic
near-threshold region.  The experimental input is given by the masses
and leptonic width of the $\Upsilon$ resonances which are known with
high accuracy. On the theoretical side the nonperturbative effects are
well under control.  This makes the $\Upsilon$ sum rules one of the
most accurate sources for the bottom quark mass value.  The complete
perturbative analysis has been performed up to
NNLO~\cite{KPP,PenPiv1,MelYel2,BenSig,Hoa1}.  The extension to 
N$^3$LO is a challenging problem.  

The theoretical value of the high moments is saturated by the
contribution of a few lowest heavy quarkonium states and the corrections
to the moments are dominated by the corrections to their masses and wave
functions at the origin. To estimate the size of the N$^3$LO corrections
we include the ${\cal O}(m_q\alpha_s^5)$ result for the energy levels
and the partial ${\cal O}(\alpha_s^3)$ result for the wave function at
the origin which includes all the logarithmic term \cite{KPSS2} and the
$\beta_0^3$ terms obtained in Section~\ref{sec2}.  We perform the
analysis along the lines described in Ref.~\cite{PenPiv1} using
$\mu=\mu_s$.  For $n\ge 20$ the corrections to the moments are dominated
by the one to the ground state energy and we recover the result of
Ref.~\cite{PenSte} for the bottom quark mass. For lower moments, which
provide better balance between theoretical end experimental
uncertainties \cite{PenPiv1}, the situation changes drastically as the
corrections to the wave function at the origin begin to play an
important role. For $n=4$ the negative third-order contribution to the
wave function completely cancels the effect of the third-order
correction to the binding energy, and the correction to the pole mass
$m_b$ almost vanishes. The pole mass can be converted into the
$\overline{\rm MS}$ mass $\bar m_b(\bar m_b)$ which is widely believed
to have much better perturbative properties. If we correlate the series
so that the $k^{\rm th}$-order correction to the sum rules goes along
with the $k$-loop mass relation, which is natural for low moments, we
obtain as an effect of the third-order corrections
$\delta\bar m_b(\bar m_b)_{\rm N^3LO}\approx -100$~MeV.  We take
this variation as an estimate for the size of the N$^3$LO corrections
within the $\Upsilon$ sum-rule approach.  It is interesting to note that
the N$^3$LO correction to $\bar m_b(\bar m_b)$ is {\it negative} at the
soft normalization scale in contrast to the series obtained from the
ground state energy analysis \cite{PenSte}.
 

\vspace*{1em}
\noindent
{\bf Top quark-antiquark  threshold production.}
The nonperturbative effects in the case of the top quark are negligible.
However, due to the relatively large top quark width, $\Gamma_t$,
its effect has to be taken into account properly~\cite{FadKho}
since the Coulomb-like resonances below threshold are smeared out.
Actually, the cross section only shows a small bump which is
essentially the remnant of the ground state pole.
The higher poles and
continuum, however, affect the position of the resonance peak and move
it to higher energy. The value of the normalized cross section
$R=\sigma(e^+e^-\to t\bar t)/\sigma(e^+e^-\to\mu^+\mu^-)$ at the
resonance energy is dominated by the contribution from the {\it  would-be} 
toponium ground state which in the leading approximation
reads $R_1^{\rm LO}=6\pi
N_cQ_t^2|\psi_1^C(0)|^2/\left(m_t^2\Gamma_t\right)$, where $Q_t=2/3$.
The analog to Eq.~(\ref{eq::gamser}) reads
\begin{eqnarray}
  R_1   &\approx& R_1^{\rm LO}
  \left(1
  -1.70\,\alpha_s(m_t)
  -7.89\,\alpha_s^2(m_t)+\ldots\right)
  \nonumber\\
  &&{}\times\left(1
  -0.43\,\alpha_s
  -5.19\,\alpha_s^2\ln\alpha_s
  +16.1\,\alpha_s^2
  \right. \nonumber\\&&{}\left.
  -13.8\,\alpha_s^3\ln^2\alpha_s
  +2.06\,\alpha_s^3\ln\alpha_s
  -27.2\,\alpha_s^3|_{\beta_0^3}
  +\ldots\right)
  \,,
  \label{rser}
\end{eqnarray}
with $\alpha_s=\alpha_s(\mu_s)$. Numerically we find
\begin{eqnarray}
  R_1 &\approx& R_1^{\rm LO}(1 -0.243_{\rm NLO}
  +0.435_{\rm NNLO}-0.268_{\rm N^3LO^\prime}+\ldots)
  \,.
  \label{rnum}
\end{eqnarray}
The new third-order corrections proportional to $\beta_0^3$ amount to
approximately $-7\%$ of the LO approximation at the soft scale which is
the same order of magnitude as the ${\cal O}(\alpha_s^3)$ linear
logarithmic term.  The available N$^3$LO terms improve the stability of
the result with respect to the scale variation as can be seen in
Fig.~\ref{fig::mu}(b).  The absence of a rapid growth of the
coefficients along with the alternating-sign character of the series and
the weak scale dependence suggest that the missing perturbative
corrections are moderate and most likely are in the few-percent range.  It
is interesting to note that the perturbative contributions of different
orders, which are relatively large when taken separately, cancel in the
sum to give only a few percent variation of the leading order result.
  


\section{Summary}
\label{sec4}
In this paper the important class of the third-order corrections to the
heavy quarkonium parameters proportional to $\beta_0^3$ has been
obtained. The complete result for the exited states spectrum to ${\cal
  O}(m_q\alpha_s^5)$ is derived. The perturbative results are in
surprisingly good agreement with the $\Upsilon(2S)$ and $\Upsilon(3S)$
meson masses and the leptonic width of the $\Upsilon(1S)$ meson. Thus the
nonperturbative effects in bottomonium seem to be rather moderate and
the theoretical results are dominated by the perturbative contributions.  A
failure of early low-order perturbative analysis to describe the
$\Upsilon$ system is due to large perturbative corrections to the
Coulomb approximation. On the basis of our results the magnitude of the
N$^3$LO corrections to the $\Upsilon$ sum rules and top quark-antiquark
threshold production cross section is estimated.  The available N$^3$LO
corrections which include all logarithmic terms and the nonlogarithmic
$\beta_0^3$ contribution stabilize the perturbative series for the
production/annihilation rates that makes us more optimistic about
possible accurate perturbative description of these quantities.


\vspace*{1em}

\noindent
{\bf Acknowledgements}\\
We thank M. Beneke, Y. Kiyo and K. Schuller for cross-checking the 
large-$\beta_0$ results prior to publication~\cite{BKS}.
A.A.P. would like to thank Y. Sumino and the theory group of Tohoku
University for hospitality.  
The work of V.A.S.
was supported in part by 
DFG Mercator Visiting Professorship No. Ha 202/1 and
Volkswagen Foundation Contract No. I/77788.
This work was supported by BMBF Grant No.\ 05HT4VKA/3, SFB/TR 9
and by the ``Impuls- und Vernetzungsfonds'' of the
Helmholtz Assciation, contract number VH-NG-008.


\begin{appendix}


\section{\label{app::coef}Static potential and beta-function}

For convenience of the reader we list in this appendix the result
for the coefficients of the static potential (see \cite{Pet,Sch,KPSS}
and references therein) 
\begin{eqnarray}
  a_1 &=& {31\over 9}C_A-{20\over 9}T_Fn_l
  \,, \nonumber\\
  a_2&=&
  \left[{4343\over162}+4\pi^2-{\pi^4\over4}+{22\over3}\zeta(3)\right]C_A^2
  -\left[{1798\over81}+{56\over3}\zeta(3)\right]C_AT_Fn_l
  \nonumber\\
  &&{}-\left[{55\over3}-16\zeta(3)\right]C_FT_Fn_l
  +\left({20\over9}T_Fn_l\right)^2
  \,,
\end{eqnarray}
and the beta-function
\begin{eqnarray}
  \beta_0&=&{1\over4}\left({11\over3}C_A-{4\over3}T_Fn_l\right)
  \,, \nonumber\\
  \beta_1&=&{1\over16}\left({34\over3}C_A^2-{20\over3}C_AT_Fn_l-4C_FT_Fn_l
  \right)
  \,, \nonumber\\
  \beta_2&=&{1\over64}\left({2857\over54}C_A^3-{1415\over27}C_A^2T_Fn_l
  -{205\over9}C_AC_FT_Fn_l+2C_F^2T_Fn_l+{158\over27}C_AT_F^2n_l^2
  \right.\nonumber\\
  &&{}+\left.{44\over9}C_FT_F^2n_l^2\right)
  \,,
\end{eqnarray}
where $T_F=1/2$ and $n_l$ is the number of the light quark flavours.


\section{\label{app::results}Results for $\delta E_n^{(i)}$}

In this appendix we collect the known results for the perturbative
corrections to the heavy quarkonium spectrum. The first and the second order
corrections read \cite{PinYnd,PenPiv1,MelYel1}
\begin{eqnarray}
  \delta E_n^{(1)} &=&
  E_n^C \frac{\alpha_s}{\pi}
  \left[
  4\beta_0\left(\lmu{n} + \PolyEuler{n+1}\right)
  +\frac{a_1}{2} 
  \right]
  \,,
  \label{eq::En1}
  \\
  \delta E_n^{(2)} &=&
  E_n^C \left(\frac{\alpha_s}{\pi}\right)^2
  \Bigg[
    12\beta_0^2\lmu{n}^2
    + \left(3 a_1\beta_0 + 4\beta_1 
    + \left(-8 + 24\PolyEuler{n+1}\right)\beta_0^2
    \right) \lmu{n} + \frac{a_1^2}{16} + \frac{a_2}{8}
    \nonumber\\&&\mbox{}
    +\left(-1 + 3\PolyEuler{n+1}\right) a_1\beta_0
    +\left(
    \frac{8}{n^2} + \frac{10\pi^2}{3} 
    - \left(8 + \frac{8}{n}\right) \PolyEuler{n+1}
    + 12\PolyEuler{n+1}^2 
    \right.\nonumber\\&&\mbox{}
     - 16\Psi_2(n) - 4 n \Psi_3(n)
    \Bigg)\beta_0^2  + 4\PolyEuler{n+1}\beta_1
    \nonumber\\&&\mbox{}
    + \frac{\pi^2}{n}C_AC_F
    + \left(\frac{2}{n}-\frac{11}{16n^2}
    -\frac{2}{3n}\sss \right) \pi^2C_F^2
  \Bigg]
  \,.
  \label{eq::En2}
\end{eqnarray}
The result for
$\delta E_n^{(3)}\Big|_{\beta(\alpha_s)=0}$ reads~\cite{KPSS1}
\begin{eqnarray}
\lefteqn{
  \left.\delta E^{(3)}_n\right|_{\beta(\alpha_s)=0}
  =}
\nonumber\\&&
-E^C_n{\alpha_s^3\over\pi}\left\{
-{a_1a_2+a_3\over32\pi^2}
+\left[-{C_AC_F\over2}+\left(-\frac{7}{4}+{9\over16n}
+{S(S+1)\over2}\right)C_F^2\right]{a_1\over n}\right.
\nonumber\\
&&{}+\left[{5\over36}+{1\over6}
\left(\ln2-\gamma_E-\ln n-\Psi_1(n+1)+L_{\alpha_s}\right)\right]C_A^3
\nonumber\\
&&{}+\left[-{97\over36}
+{4\over3}\left(\ln2+\gamma_E-\ln n+\Psi_1(n+1)+L_{\alpha_s}\right)
\right]{C_A^2C_F\over n}
\nonumber\\
&&{}+\left[\left(-{139\over36}+4\ln2+{7\over6}(\gamma_E-\ln n+\Psi_1(n+1))
+{41\over6}L_{\alpha_s}\right)\right.
\nonumber\\
&&{}+\left({47\over24}
+{2\over3}\left(-\ln2+\gamma_E+\ln n+\Psi_1(n+1)-L_{\alpha_s}\right)\right)
{1\over n}
\nonumber\\
&&{}+\left.
\left({107\over108}-{7\over12n}
+{7\over 6}(\gamma_E-\ln{n}+\Psi_1(n+1)-L_{\alpha_s})\right)S(S+1)\right]
{C_AC_F^2\over n}
\nonumber\\
&&{}+\left[{79\over18}-{7\over6n}+{8\over3}\ln2
+{7\over3}(\gamma_E-\ln n+\Psi_1(n+1))+3L_{\alpha_s}-{S(S+1)\over3}\right]
{C_F^3\over n}
\nonumber\\
&&{}+\left[-{32\over15}+2\ln2+(1-\ln2)S(S+1)\right]{C_F^2T_F\over n}
\nonumber\\
&&{}\left.+{49C_AC_FT_Fn_l\over36n}
+\left[{8\over9}-{5\over18n}-{10\over27}S(S+1)\right]{C_F^2T_Fn_l\over n}
+{2\over3}C_F^3L^E_n\right\},
\label{spectrum}
\end{eqnarray}
where $\Las=-\ln(C_F\alpha_s)$ and $\LE{n}$ stands for the QCD Bethe
logarithms with the numerical values \cite{KniPen1}
\begin{eqnarray}
\LE{1}=-81.5379\,,\qquad \LE{2}=-37.6710\,,\qquad \LE{3}=-22.4818\,.
\end{eqnarray}
The terms proportional to $\Las$ have been computed for the first time
in Ref.~\cite{BPSV1}.


\end{appendix}



\end{document}